\documentclass[preprint,3p]{elsarticle}

\usepackage[utf8]{inputenc}
\usepackage{amsmath}
\usepackage{amsfonts}
\usepackage{amssymb}
\usepackage{graphicx}
\usepackage{hyperref}
\usepackage{braket}
\usepackage{multirow}
\usepackage{enumitem}
\usepackage{bbm}
\usepackage{units}
\usepackage{ulem}

\DeclareMathOperator\tr{Tr}
\DeclareMathOperator\tTr{tTr}

\newcommand{\Qhat}{{\widehat Q}}
\def\sl3c{\text{SL}(3,\mathbb{C})}
\def\su3{\text{SU(3)}}

\graphicspath{{plots/}}

\newcommand{\comment}[1]{}

\begin{document}

\title{Improved local truncation schemes for the higher-order tensor renormalization group method}

\author[1]{Jacques Bloch}
\ead{jacques.bloch@ur.de}
\author[1,2]{Robert Lohmayer}
\ead{robert.lohmayer@ur.de}
\author[1]{Maximilian Meister}
\author[1]{Michael Nunhofer}

\address[1]{Institute for Theoretical Physics, University of Regensburg, 93040
  Regensburg, Germany}
\address[2]{Leibniz Institute for Immunotherapy (LIT), 93053 Regensburg, Germany}

\date{October 5, 2022}

\begin{abstract}
The higher-order tensor renormalization group is a tensor-network method providing estimates for the partition function and thermodynamical observables of classical and quantum systems in thermal equilibrium. At every step of the iterative blocking procedure, the coarse-grid tensor is truncated to keep the tensor dimension under control. For a consistent tensor blocking procedure, it is crucial that the forward and backward tensor modes are projected on the same lower dimensional subspaces. In this paper we present two methods, the SuperQ and the iterative SuperQ method, to construct tensor truncations that reduce or even minimize the local approximation errors, while satisfying this constraint.  
\end{abstract}

\maketitle

\section{Introduction}
\label{Sec:Intro}

Physical systems in thermal equilibrium are described by their partition function, whose complexity grows exponentially in the volume. The standard method to simulate such statistical systems is the Markov chain Monte Carlo method (MC), which efficiently samples the relevant states of the system to produce reliable estimates of observables. A fundamental prerequisite for the MC method is the positivity of the sampling weights. Models which do not satisfy this condition cause the infamous sign problem and require alternative simulation methods. Quantum systems with complex actions are typical examples of systems with a sign problem. An important topical application in high energy physics is the simulation of quantum chromodynamics (QCD) at nonzero quark chemical potential, which allows for the investigation of the QCD phase diagram as a function of temperature and baryon density.

There exist numerous methods to circumvent the sign problem, and some even solve it for particular systems \cite{Aarts:2015kea,Aarts:2014fsa,Gattringer:2016kco}. Very mild sign problems can be circumvented by reweighting, which uses the Monte Carlo method on an auxiliary ensemble with positive weights, and reweights the observables to the target ensemble. The main issue with this method is that the statistical error increases exponentially with the volume such that it is hardly usable in any realistic situation, except for the validation of other methods in regions where the sign problem is small. Other methods which have shown their merit on some models, but are known to have fundamental problems for other ones, are the complex Langevin method, the thimbles, the density of states method and the method of dual variables, where the simulations are usually performed with the worm algorithm.  Common to those methods is the stochastic sampling of the partition function. 

An alternative approach that has recently drawn a lot of interest is that of  tensor networks, see \cite{Meurice:2020pxc} for a review. In these methods the partition function is first rewritten as a full contraction of a tensor network covering the entire lattice. The exact computation of the partition function and observables in this formulation would have an exponential complexity. 
The tensor renormalization group (TRG) \cite{Levin:2006jai} and higher order tensor renormalization group (HOTRG) \cite{Xie_2012} methods avoid this exponential cost by blocking the lattice iteratively and truncating the inflated dimensions of the coarse grid tensor at each blocking step using truncated higher order singular value decompositions (HOSVD) \cite{DeLathauwer2000}, which are based on the matrix singular value decomposition (SVD).

We consider the partition function of a  $d$-dimensional classical or quantum system in thermal equilibrium, written as a fully contracted tensor network \cite{Liu:2013nsa},
\begin{align}
Z = \tTr \prod_{x=1}^V T^{(x)}_{i_{x,-1} i_{x,1} \dots i_{x,-d} i_{x,d}} ,
\end{align}
with a tensor $T^{(x)}$ at each site $x=1,\ldots,V$. In general the local tensor is the same on all sites, i.e.,
$T^{(x)}=T$ for all $x$. 
For each lattice direction $\nu$, the tensor has one mode for the forward and one mode for the backward orientation, corresponding to the indices $i_{x,\nu}$ and $i_{x,-\nu}\equiv i_{x-\hat\nu,\nu}$, respectively, where  $\hat\nu$ is a unit step in the $\nu$ direction.
We will often refer to these modes as backward and forward modes of the physical tensor.
The trace in the partition function stands for a full contraction over all tensor indices, where two adjacent tensors share exactly one index. 

Thermodynamical observables, which are defined as derivatives of the partition function with respect to one of its parameters, can be computed using either a finite-difference approximation or an impurity tensor formulation involving the analytical derivative of $T$ \cite{Zhao_2016}.

\begin{figure}
\centering
\includegraphics[scale=1]{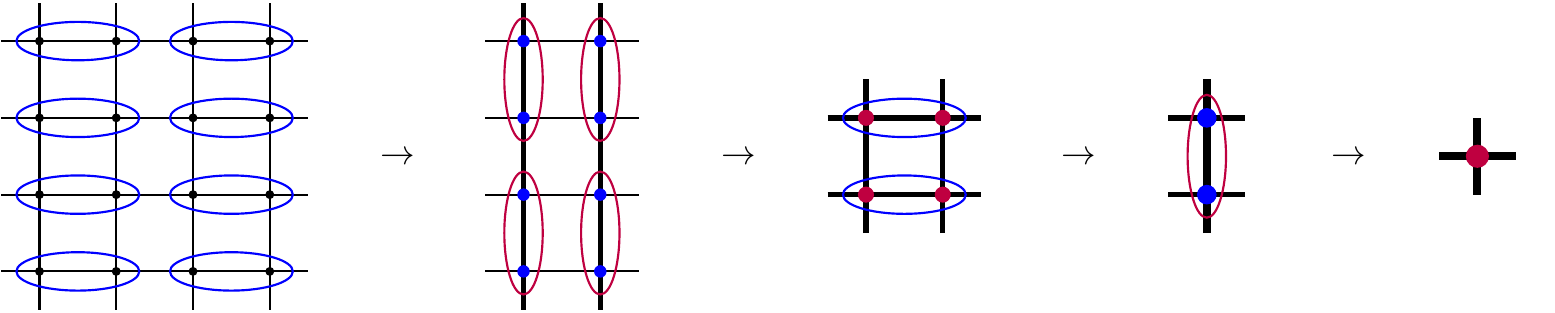}
\caption{Blocking procedure to reduce a two-dimensional $4\times 4$ lattice to a single tensor using alternating contractions in the horizontal and vertical directions.}
\label{Fig:RGblocking}
\end{figure}

In the following we will restrict our discussion to HOTRG, because it can be applied to any number of dimensions, whereas TRG is limited to two-dimensional systems. 
The HOTRG method uses an iterative blocking procedure that reduces the size of the lattice by a factor of two during each blocking step by contracting pairs of adjacent tensors. 
The procedure is illustrated for a two-dimensional $4\times 4$ lattice in Fig.~\ref{Fig:RGblocking}. Its extension to higher dimensions is obvious, and below we will further discuss the HOTRG method for the three-dimensional case.

\begin{figure}
\centering
\includegraphics[scale=3.2]{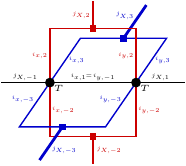}
\caption{Illustration of the contraction $T\star_1 T=M$ along the $1$-direction in a three-dimensional system, as in \eqref{Mhotrg}. The square nodes represent the fusion of the original tensor indices into the combined fat indices \eqref{fatmodes} of $M$.}
\label{contraction}
\end{figure}

When contracting two adjacent tensors $T$ over their shared link, a tensor $M$ of higher order is produced. Such a contraction in the $1$-direction is illustrated for the three-dimensional case in Fig.~\ref{contraction} and can be written as
\begin{align}
&M_{j_{X,{-1}}j_{X,1}j_{X,-2}j_{X,2}j_{X,-3}j_{X,3}} = \sum\limits_{i_{x,1}} 
T_{i_{x,-1} i_{x,1}i_{x,-2} i_{x,2}i_{x,-3} i_{x,3}} \, T_{i_{y,-1} i_{y,1}i_{y,-2} i_{y,2}i_{y,-3} y_{x,3}} 
\label{Mhotrg}
\end{align}
where $y = x+\hat1$ and therefore $i_{y,-1}=i_{x,1}$, by definition, $X=(x,y)$ labels sites on the coarse grid and
\begin{align}
&\begin{aligned}
&j_{X,-1}=i_{x,-1},  \qquad\qquad\, j_{X,1}=i_{y,1} \\
&\begin{aligned}
&j_{X,-2}=(i_{x,-2},i_{y,-2}), && j_{X,2}=(i_{x,2},i_{y,2}) \\
&j_{X,-3}=(i_{x,-3},i_{y,-3}), && j_{X,3}=(i_{x,3},i_{y,3}) 
\end{aligned}
\quad\Bigg\}
\quad\text{\textbf{fat} indices}.
\end{aligned}
\label{fatmodes}
\end{align}
For any direction perpendicular to the direction of contraction, the tensor $M$ has modes originating from both contracted tensors. 
To keep the order of the tensor unchanged, we gather every such pair of modes in a new fat mode corresponding to its direct product space. Assuming that the modes of the local tensor have dimension $D$, then the fat modes will have dimension $D^2$. In HOTRG these fat modes are truncated back to dimension $D$ using a modified version of the HOSVD approximation method, such that the dimension of the coarse grid tensor remains the same as that of the original local tensor throughout the entire blocking procedure. 

In general, step $k+1$ of the HOTRG procedure can be summarized as
\begin{align}
T^{[k]} \star_\nu T^{[k]} =: M \stackrel{\text{truncate}}{\longrightarrow} T^{[k+1]} ,
\label{T(k+1)}
\end{align}
where the $\star_\nu$-operation symbolically represents a forward-backward contraction in direction $\nu$. The precise construction of $T^{[k+1]}$ will be discussed in Sec.~\ref{Sec:BFcond}.

In the standard approximation procedure using HOSVD \cite{DeLathauwer2000}, referred to as \textit{HOSVD approximation} in the following, the dimension of each tensor mode gets reduced by projecting it on a lower dimensional subspace, which is generically different for each mode.
This HOSVD approximation is modified when used as part of the iterative blocking procedure in the standard HOTRG algorithm, as it is essential for the accuracy and effectiveness of the method that the backward and forward modes for every direction get projected on the same subspace. 
Each of these subspaces will be characterized by a frame, which is a set of orthonormal basis vectors spanning the subspace. Constructing appropriate frames will be the major subject of this paper. 

The standard HOTRG procedure for the construction of frames \cite{Xie_2012} is not optimal, in particular when the local tensor is not symmetric in its backward and forward modes. In this paper we present two improved methods for the construction of common subspaces for pairs of backward and forward modes: 
the \textit{SuperQ} and the \textit{iterative SuperQ} method (ISQ), which is an iterative improvement of the former in search of the optimal subspaces. 
Note that the discussion in this paper solely focusses on the optimization of the rank reduction of the local tensors at every blocking step, but does not take into account global effects on the full contraction of the tensor network.

Here is a brief outline of the paper. In Sec.~\ref{Sec:HOSVD} we review the standard HOSVD method to construct a reduced rank approximation for an arbitrary tensor. In Sec.~\ref{Sec:BFcond} we explain why the HOTRG method uses a modification of this rank reduction procedure such that the backward and forward modes are projected on the same subspace. We then propose two methods to improve the standard HOTRG truncation: In Sec.~\ref{Sec:SuperQ} we present the SuperQ method, and in Sec.~\ref{Sec:BFHOOI} we derive the more sophisticated ISQ method. Finally, we summarize and conclude in Sec.~\ref{Sec:conclusions}.

\section{Rank reduction and HOSVD approximation}
\label{Sec:HOSVD}

Below we first review the general idea of rank reduction for an arbitrary tensor, before describing the HOSVD procedure \cite{DeLathauwer2000} which can be used to generate a quasi-optimal rank-reduced approximation in an efficient way.
 
For a real tensor $M$ of order $n$ with dimension $N_1\times \cdots \times N_n$, the left \textit{matrix-tensor multiplication} $A \odot_r M$ is defined as a contraction of the second index of the matrix $A$ with the $r$-th index of the tensor $M$,
\begin{align}
(A \odot_r M)_{i_1\cdots i_{r-1} j_r i_{r+1} \cdots i_n} 
&= \sum_{i_r} A_{j_r i_r} M_{i_1\cdots i_n} .
\end{align}
 
A lower-rank approximation of $M$ can be constructed as
\begin{align}
\hat M &= P^{(1)} \odot_1 P^{(2)} \odot_2 \cdots P^{(n)} \odot_{n} M , 
\label{hatMP}
\end{align}
using $N_r\times N_r$ projectors $P^{(r)}$ of rank $K_r\leq N_r$, $r=1,\dots,n$. In this approximation, the $r$-th tensor mode of dimension $N_r$ is projected onto a subspace of dimension $K_r$, embedded in the original space. 
Typically the Frobenius norm $\|M-\hat M\|$ is used as a measure for the quality of the low-rank approximation and the aim is to determine optimal projectors $P^{(r)}$ with fixed ranks $K_r$.

The projectors can be represented as
\begin{align}
P^{(r)} = U^{(r)} U^{(r)^T} 
\end{align}
with semi-orthogonal $N_r\times K_r$  matrices $U^{(r)}$, which we will call \textit{frames} in the following. 
Semi-orthogonal means that the columns of the frames are orthonormal, but not their rows (unless $K_r=N_r$).
The columns of each frame $U^{(r)}$ provide an orthonormal basis of the corresponding $K_r$-dimensional subspace,
\begin{align}
U^{(r)^T}U^{(r)} = \mathbbm{1}_{K_r} .
\end{align}

The approximation $\hat M$ can then be rewritten as
\begin{align}
\hat M = U^{(1)} \odot_1 U^{(2)} \odot_2 \cdots U^{(n)} \odot_{n} S  ,
\label{hatM}
\end{align}
with the $K_1\times\dots\times K_n$ dimensional \textit{core tensor}
\begin{align}
S = U^{(1)^T} \odot_1 U^{(2)^T} \odot_2 \cdots U^{(n)^T} \odot_{n} M .
\label{hatS}
\end{align}
The core tensor represents $\hat M$ in the bases of the subspaces, which are spanned by the columns of the frames $U^{(r)}$.
Note that $\hat M$ has the same dimension as $M$, but is generically of lower rank as the $r$-th mode is projected from a space of dimension $N_r$ on a subspace of dimension $K_r$ for $r=1,\ldots,n$. 
In practice, $\hat M$ is usually not constructed explicitly, as many operations involving $\hat M$ can be performed at much lower cost using only the core tensor $S$ and the frames $U^{(r)}$, which contain the same information as $\hat M$ but condensed in lower-dimensional objects. A typical example of such an operation is the contraction of two tensors, as will be discussed in the next section.

The squared Frobenius norm of $M$ is given by
\begin{align}
\| M \|^2 = \braket{M,M} = \sum_{\{i\}} M_{i_1\cdots i_n}^2 ,
\label{Frobnorm}
\end{align}
where for brevity we introduce the notation $\{i\}={i_1,\dots,i_{n}}$ for the summation indices, and the inner product between two real tensors of equal dimension is defined as
\begin{align}
\braket{A,B} = \sum_{\{i\}} A_{i_1\cdots i_n} B_{i_1\cdots i_n} .
\end{align}

Since the reduced-rank tensor $\hat M$ is a projection of $M$, we have
\begin{align}
\braket{M,\hat M} = \|\hat M\|^2 =  \|S\|^2 .
\end{align}
Therefore, the squared approximation error is given by
\begin{align}
\|M-\hat M\|^2 = \|M\|^2 + \|\hat M\|^2 - 2 \braket{M,\hat M}
= \|M\|^2 - \|\hat M\|^2 = \|M\|^2 - \|S\|^2  ,
\label{normapprox}
\end{align}
which does not require the explicit computation of $\hat M$.

In the HOSVD approximation procedure \cite{DeLathauwer2000} the semi-orthogonal frames used to approximate $M$ are constructed using properties of matrix SVDs. We first introduce the $r$-unfolding $M^{(r)}$, which is a matrix containing the same entries as the tensor $M$, but reordered such that its rows correspond to the $r$-th mode of $M$ and its columns correspond to a combination of all other tensor modes. The entries of the $r$-unfolding of $M$ are given by
\begin{align}
M^{(r)}_{i_r\,,\,(\{i\}\setminus i_r)} = M_{i_1\cdots i_n} ,
\label{unfold}
\end{align}
where the column index $(\{i\}\setminus i_r) = (i_1,\dots,i_{r-1},i_{r+1},\dots,i_n)$ is a linear index of coordinates in a space of dimension $\prod_{s\neq r} N_s$. 
The multi-rank of a tensor is defined by the $n$-tuple of the ranks of the individual unfoldings $M^{(r)}$, $r=1,\ldots,n$. Therefore $M$ has at most multi-rank $(N_1,\ldots,N_n)$ and the approximation $\hat M$ of \eqref{hatMP} at most multi-rank $(K_1,\ldots,K_n)$.
Note that the squared Frobenius norm \eqref{Frobnorm} of a tensor is identical to that of any of its unfoldings, as it is just a sum over all squared components,
\begin{align}
\| M \|^2 = \| M^{(r)} \|^2 = \tr \left[ M^{(r)} M^{(r)^T} \right]
\label{FrobnormGramian}
\end{align}
for any $r=1,\ldots,n$.
To construct the HOSVD approximation $\hat M$, we first consider the singular value decomposition for each $r$-unfolding $M^{(r)}$ of $M$ (assuming real tensors for simplicity),
\begin{align}
M^{(r)} = L^{(r)} \Sigma^{(r)} R^{(r)^T} ,
\label{SVD}
\end{align}
where the columns of the $N_r \times N_r$ orthogonal matrix $L^{(r)}$ are the left singular vectors of $M^{(r)}$. The columns of the orthogonal matrix $R^{(r)}$ contain the corresponding right singular vectors of $M^{(r)}$.
The diagonal entries of $\Sigma^{(r)}$ are the singular values of $M^{(r)}$, which are always real and non-negative, while all other entries are zero. 

It is well-known in linear algebra that retaining the $K_r$ largest singular values in $\Sigma^{(r)}$, while setting all others to zero, yields the best-possible matrix approximation $A^{(r)}$ of rank $K_r$ to $M^{(r)}$ (best-possible referring to a minimization of the Frobenius norm $\|M^{(r)}-A^{(r)}\|$).
The relative truncation error is given by
\begin{align}
\epsilon^{(r)} = \frac{\|M^{(r)}-A^{(r)}\|}{\|M^{(r)}\|} = \sqrt{\frac{\sum_{i=K_r+1}^{N_r} \lambda^{(r)}_i}{\sum_{i=1}^{N_r} \lambda^{(r)}_i}} , 
\label{trerr}
\end{align}
where $\lambda^{(r)}_i$ are the eigenvalues of the Gramian $M^{(r)} M^{(r)^T}$, i.e., the squared singular values of the unfolding $M^{(r)}$, ordered such that $\lambda^{(r)}_1 \geq \dots \geq \lambda^{(r)}_{N_r}$.

This matrix property is used in HOSVD by separately performing the matrix SVDs of all individual unfoldings $M^{(r)}$ of $M$ for $r=1,\ldots,n$  and constructing the $N_r\times K_r$ frames $U^{(r)}$ with the singular vectors of $L^{(r)}$ corresponding to the $K_r$  largest singular values of the unfoldings $M^{(r)}$. 
These frames are then used to construct the core tensor \eqref{hatS} and the matrix approximation \eqref{hatM} of HOSVD.
Unlike for the matrix case, the HOSVD tensor approximation is in general not the best-possible approximation of a given multi-rank $(K_1,\ldots,K_n)$, even though it is usually quite close to it \cite{DeLathauwer2000}.

In a variant of the HOSVD approximation, called interlaced HOSVD approximation, the rank reduction procedure is carried out in the following way: Starting with $S=M$, the frames are computed on successive unfoldings of the core tensor, which gets updated every time a new truncation frame is constructed until the core tensor is of dimension $K_1\times\ldots\times K_n$.
For the interlaced HOSVD approximation, the result depends on the order of the updates, but is usually close to that of the ordinary HOSVD approximation.

The best-possible approximation of multi-rank $(K_1,\ldots,K_n)$, which minimizes the Frobenius norm $\|M-\hat M\|$, can be constructed numerically using the Higher Order Orthogonal Iteration (HOOI) \cite{DeLathauwer2000a}. 
Nevertheless, the HOSVD approximation is especially appealing because of its relative simplicity to produce an almost optimal approximation.

\section{Backward-forward symmetric truncation in HOTRG}
\label{Sec:BFcond}

We now discuss how the HOSVD formalism is used in HOTRG to avoid the exponential blow up of the tensor dimension during the blocking procedure, and why the standard HOSVD truncation is modified to avoid drawbacks related to accuracy and efficiency. To make our point we will use the two-dimensional case as it can be easiest illustrated and contains all the ingredients necessary for the discussion. Extending it to higher dimensions is straightforward.

\begin{figure}
\centering
\includegraphics{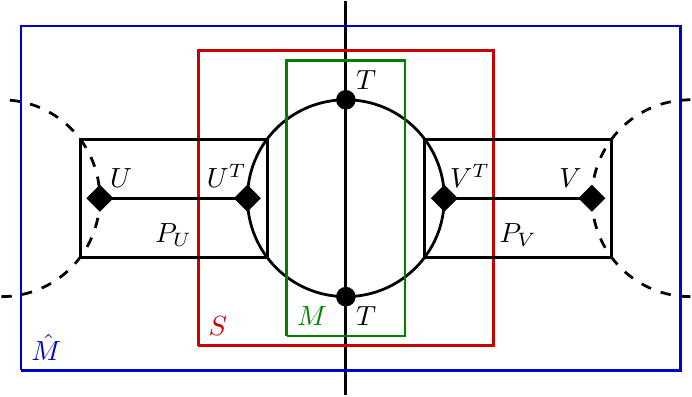}
\caption{Two tensors $T$ are contracted over their shared vertical link, producing a tensor $M$ with fat horizontal modes (green box). The fat modes of $M$ are projected onto subspaces with the projectors $P_U=U U^T$ and $P_V=V V^T$, respectively, to form the lower rank approximation $\hat M$ of \eqref{hatM-hotrg} (blue box). As part of the construction one recognizes the core tensor $S$ of \eqref{core} (red box). Note that the matrices $U$, $U^T$, $V$ and $V^T$,  described by diamonds in the figure, are applied from the inside to the outside, in correspondence with \eqref{core} and \eqref{hatM-hotrg}.}
\label{Fig:tildeM}
\end{figure}

We consider the contraction $M=T\star_2 T$ of two local tensors $T$ along the $2$-direction.
According to the discussion in the introduction, $M$ will have thin backward and forward modes of dimension $D$ in the contracted $2$-direction, and fat backward and forward modes of dimension $D^2$ in the perpendicular $1$-direction, which we want to reduce to lower rank by projecting on a $D$-dimensional subspace using \eqref{hatMP}.
This procedure of contraction and truncation, which we detail below, is illustrated in Fig.\ \ref{Fig:tildeM}. 
To reduce the dimension of the fat modes back from $D^2$ to $D$, while minimizing the loss of information, we apply the HOSVD approximation procedure, explained in Sec.~\ref{Sec:HOSVD}, where we only truncate the fat modes. 
The SVDs are computed for the unfoldings $M^{(1)}$ and $M^{(2)}$ for the backward and forward modes in the 1-direction, respectively, and the frames $U$ and $V$ of dimension $D^2\times D$ are constructed with the singular vectors corresponding to their $D$ largest singular values. For the modes in the contracted $2$-direction no truncation is required.
With these frames we construct a core tensor $S$ of dimension $D \times D \times D \times D$, according to \eqref{hatS},
\begin{align}
S = U^T \odot_{1} V^T \odot_2 M .
\label{core}
\end{align}
The corresponding approximation $\hat M$, defined in \eqref{hatM}, with the same dimension as $M$, but typically much lower rank, is given by
\begin{align}
\hat M = U \odot_1 V \odot_2 S
= P_U \odot_1 P_V \odot_2 M ,
\label{hatM-hotrg}
\end{align}
with $D^2\times D^2$ projectors $P_U = U U^T$ and $P_V = V V^T$, with $U^T U = V^T V = \mathbbm{1}_D$. 
As mentioned in Sec.\ \ref{Sec:HOSVD}, 
operations involving $\hat M$ can typically be performed at much lower cost by using only the core tensor $S$ and the frames $U$ and $V$.

\begin{figure}
\centering
\includegraphics{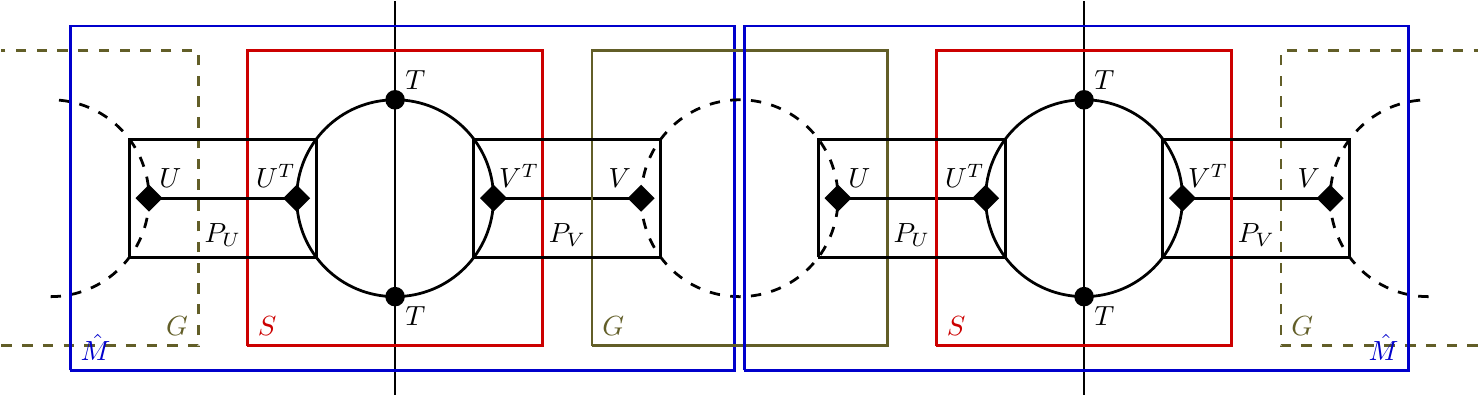}
\caption{Two approximations $\hat M$, constructed in Fig.~\ref{Fig:tildeM}, are contracted in the horizontal direction. This illustrates how the projections performed in the first contraction are concatenated when making this second contraction, leading to a product $P_U P_V$. Here a new building block $G=U^T V$ arises, which we call a \textit{merger} between two core tensors $S$. Note that, for consistency, the matrices operate in chronological order (from the inside to the outside with respect to $M$ of Fig.\ \ref{Fig:tildeM}) and not from left to right. The half-mergers on the left and right will connect to their counter parts in further contractions.}
\label{Fig:truncation}
\end{figure}

Assume that in the next blocking step two $\hat M$ tensors are contracted in the $1$-direction,  as is illustrated in Fig.~\ref{Fig:truncation}. When using the standard HOSVD approximation \eqref{hatM-hotrg} the backward and forward modes in $\hat M$ will have been projected on different subspaces using the projectors $P_U$ and $P_V$, respectively. 
In a contraction $\hat M \star_1 \hat M$ the two projectors will be multiplied, as can be seen in the center of the figure. 
The decomposition of the approximation $\hat M$ in \eqref{hatM-hotrg} can be used to reduce the computational effort, as the original contraction can be replaced by contractions of two core tensors $S$ with a $D\times D$ dimensional merger $G \equiv U^T V$ in between, as can be seen in the figure.\footnote{In fact this produces an amputated version of $\hat M \star_1 \hat M$, which together with the mergers is all we need in the full contraction of the tensor network, see also Fig.\ \ref{Fig:truncation}.} The entries of $G$ are scalar products of the basis vectors in $U$ and $V$.

At this point it is important to discuss a crucial modification introduced by the HOTRG method to the HOSVD truncation procedure presented above, which is rarely discussed in the literature.
Although the HOSVD approximation $\hat M$ gives a close-to-best lower-rank approximation to $M$, it is in general not such a good and useful truncation when viewed as part of the iterative blocking procedure. 
Indeed, the product of projectors corresponds to a projection of a projection, which will unavoidably loose additional information if the projectors $P_U$ and $P_V$ are different. In this case, the  contraction $\hat M \star_1 \hat M$ will no longer necessarily be a good approximation of $M \star_1 M$, even if $\hat M$ itself was close to the best-possible approximation of $M$. 

\begin{figure}
\centering
\includegraphics{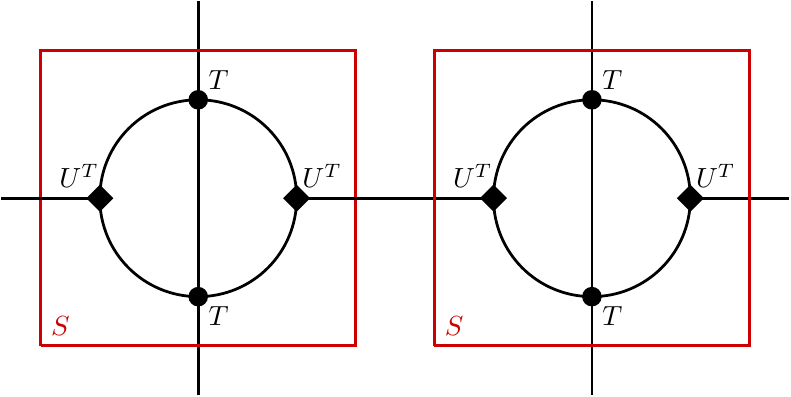}
\caption{The construction in Fig.\ \ref{Fig:truncation} drastically simplifies when choosing $U=V$, as the merger $G=\mathbbm{1}_D$ in this case. The frames are solely needed to construct the core tensor $S$. This property will spread through the entire blocking procedure (including the final trace), such that the calculation can be performed using the core tensor only.}
\label{Fig:truncation_sym}
\end{figure}

We now observe that, due to the idempotence of projectors, there would be no additional loss if $P_U=P_V$, i.e., if the backward and forward modes of $M$ in the 1-direction were projected on the same subspace. Note that in this case the merger $G$ is an orthogonal matrix. Moreover, we can also get a serious gain in algorithmic simplicity, on top of this accuracy improvement, if we choose the same basis for both modes in the common subspace, i.e., we choose frames satisfying $U=V$, for which $G=\mathbbm 1_D$. When looking back at Fig.~\ref{Fig:truncation} we see that, in this case, the central merger just drops out, and the contraction can be replaced by a contraction of two core tensors, as is illustrated in Fig.~\ref{Fig:truncation_sym} (the frames on the left and right of Fig.\ \ref{Fig:truncation} will connect to their counter parts in further contractions, to form another merger $G=\mathbbm 1_D$, which will also drop out). 

For this reason, in HOTRG the backward and forward modes of each direction are truncated using a common frame $U^{(\nu)}$, $\nu=1,\dots,d$, even when the HOSVD frames are different, which is typically the case for systems at nonzero chemical potential. With this backward-forward symmetric truncation, the core tensor $S$ can be used as new coarse grid tensor after each blocking step, where at the $(k+1)$-th blocking step two tensors of step $k$ are contracted to form a new coarse grid tensor,
\begin{align}
T^{[k]} \star_\nu T^{[k]} =: M \stackrel{\text{BF}}{\longrightarrow} S =: T^{[k+1]} .
\label{T(k+1)-specific}
\end{align}
The abbreviation BF on the arrow means that we apply a backward-forward symmetric truncation to construct the core tensor, which then becomes the new local tensor on the coarse grid.
The frames are only needed to construct the core tensor with \eqref{core}, and can then be discarded.

The same reasoning also holds for a contraction in the other direction, where the directions of thin and fat modes are interchanged. Moreover, the procedure naturally generalizes to $d$ dimensions, where we have $d-1$ directions $\nu$ with backward and forward fat modes: If the backward and forward frames are chosen to be identical for each direction, the core tensor \eqref{hatS} can be used as the new coarse grid tensor in the HOTRG blocking procedure.

This strategy of choosing the same frame to truncate the backward and forward mode for each individual direction in HOTRG, makes it fundamentally different from the HOSVD approximation, as it can no longer directly rely on the optimal low-rank properties of matrix SVD.
Geometrically, truncating the backward and forward modes with the same frame means that these modes get projected on the same subspace and are described in the same basis. This explains why the full contraction of the tensor network into a scalar can be rewritten in terms of the core tensors only.

Note that if one would use the standard HOSVD approximation procedure and work with different backward and forward frames, we would have to use both the core tensors and the mergers $G$ defined above when performing the iterative contractions, to ensure that the different subspaces are matched onto one another, see Fig.~\ref{Fig:truncation}. Although this is no conceptual problem, it would complicate the algorithm, require additional computational work, and most of all the product of projectors would deteriorate the results further.

The construction of the shared backward-forward frames in HOTRG has not been given a lot of attention in the literature until now. There is a brief discussion of this issue in the original HOTRG paper \cite{Xie_2012}, where either the backward or forward frame is chosen and applied to both modes, depending on which one gives the smallest SVD truncation error. The error introduced by this choice on the other mode is however not taken into account. We observed that for tensors lacking a backward-forward symmetry, this choice of frame is not optimal and can be improved upon.

Below we present two new methods to improve the construction of shared frames for the backward and forward modes. The first one, called SuperQ method and presented in Sec.~\ref{Sec:SuperQ}, minimizes a combined error on the backward and forward unfoldings for each individual direction. The second method, which we call iterative SuperQ (ISQ) method is presented in Sec.\ \ref{Sec:BFHOOI}. This iterative method aims at determining the best-possible approximation to $M$ for a given multi-rank, satisfying the requirement that the backward and forward frames for each direction are identical. The ISQ method leans on ideas developed for the higher order orthogonal iteration (HOOI) method \cite{DeLathauwer2000a}, which constructs the best-possible approximation of a given multi-rank with independent frames for all modes.  We will see that the SuperQ solution can be used as a natural starting point for the ISQ procedure. Note that the SuperQ and ISQ methods are specifically conceived for tensors which are part of a physical tensor network on a space-time lattice and have modes corresponding to backward and forward orientations.

\section{The SuperQ method}
\label{Sec:SuperQ}

To discuss the construction of truncations satisfying the requirement that the frames for the backward and forward modes are identical, we consider a tensor $M$ with $d$ pairs of backward and forward modes. The tensor is thus of order $2d$ with dimension $N_1 \times N_1 \times \cdots \times N_d \times N_d$, which will be truncated to dimension $K_1 \times K_1 \times \cdots \times K_d \times K_d$ using $d$ semi-orthogonal frames $U^{(\nu)}$ of dimension $N_\nu \times K_\nu$. With these frames the core tensor is constructed using
\begin{align}
S = U^{(1)^T} \odot_1 U^{(1)^T} \odot_2 \cdots U^{(d)^T} \odot_{2d-1} U^{(d)^T} \odot_{2d} M .
\label{core-d}
\end{align}
Consider the positive semi-definite Gram matrices 
\begin{align}
Q_\text{b}^{(\nu)} =  M_\text{b}^{(\nu)} M_\text{b}^{(\nu)^T} , \qquad
Q_\text{f}^{(\nu)} =  M_\text{f}^{(\nu)} M_\text{f}^{(\nu)^T} ,
\label{BFGramians}
\end{align}
where $M_\text{b}^{(\nu)}$ and $M_\text{f}^{(\nu)}$ are the unfoldings of $M$ with respect to the backward and forward modes for direction $\nu$, i.e.,
\begin{align}
M_\text{b}^{(\nu)} \equiv M^{(2\nu-1)} , \qquad
M_\text{f}^{(\nu)} \equiv M^{(2\nu)} ,
\end{align}
and $M^{(r)}$ is the $r$-unfolding of the tensor $M$ defined in \eqref{unfold}. 
We denote the $N_\nu\times K_\nu$ frames constructed with the eigenvectors corresponding to the $K_\nu$ largest eigenvalues of $Q_\text{b}^{(\nu)}$ and $Q_\text{f}^{(\nu)}$ as $U_\text{b}^{(\nu)}$ and $U_\text{f}^{(\nu)}$, respectively.
As the backward and forward Gramians $Q_\text{b}^{(\nu)}$ and $Q_\text{f}^{(\nu)}$ are in general not identical, the corresponding subspaces spanned by the vectors of the frames $U_\text{b}^{(\nu)}$ and $U_\text{f}^{(\nu)}$ will be different too.\footnote{This can even be the case if the eigenvalues of both Gramians coincide, as we have observed for the $O(2)$ model with chemical potential.}
In the standard HOTRG procedure \cite{Xie_2012} it is suggested to choose either $U_\text{b}^{(\nu)}$ or $U_\text{f}^{(\nu)}$ for the unique $U^{(\nu)}$, depending which of both gives the smallest SVD truncation error \eqref{trerr}. Even though this choice of frame optimizes the truncation error for one mode, it does not take into account its effect on the mode corresponding to the opposite orientation. Therefore, it is clear that, generically, better choices of frames should exist, and our aim is to construct frames $U^{(\nu)}$ that reduce the combined truncation error when applied simultaneously to the backward and forward modes for the $\nu$ direction.

Let us now consider a single truncation frame $U^{(\nu)}$ which we use to reduce the rank of the unfoldings $M^{(\nu)}_\text{b}$ and $M^{(\nu)}_\text{f}$. Using \eqref{normapprox} and \eqref{FrobnormGramian}, the relative truncation errors on the backward and forward unfoldings are 
\begin{align}
\epsilon^{(\nu)}_\text{b} &= \frac{\|M^{(\nu)}_\text{b}-A^{(\nu)}_\text{b}\|}{\|M^{(\nu)}_\text{b}\|} = \sqrt{1-\frac{\tr\left(U^{(\nu)^T} Q^{(\nu)}_\text{b} U^{(\nu)}\right)}{\tr Q^{(\nu)}_\text{b}}} , \\
\epsilon^{(\nu)}_\text{f} &= \frac{\|M^{(\nu)}_\text{f}-A^{(\nu)}_\text{f}\|}{\|M^{(\nu)}_\text{f}\|} = \sqrt{1-\frac{\tr\left(U^{(\nu)^T} Q^{(\nu)}_\text{f} U^{(\nu)}\right)}{\tr Q^{(\nu)}_\text{f}}} ,
\end{align}
where $A^{(\nu)}_\text{b}=U^{(\nu)}U^{{(\nu)^T}}M^{(\nu)}_\text{b}$ and $A^{(\nu)}_\text{f}=U^{(\nu)}U^{{(\nu)^T}}M^{(\nu)}_\text{f}$ are the rank-$K_\nu$ approximations to the unfoldings $M^{(\nu)}_\text{b}$ and $M^{(\nu)}_\text{f}$, respectively, obtained with the same frame $U^{(\nu)}$.

To improve upon using either $U^{(\nu)}_\text{b}$ or $U^{(\nu)}_\text{f}$, we determine the common $U^{(\nu)}$ by minimizing the combination of both errors in 
\begin{align}
{\epsilon^{(\nu)}_\text{S}}^2 &= 
\frac12\left({\epsilon^{(\nu)}_\text{b}}^2+{\epsilon^{(\nu)}_\text{f}}^2\right) 
= 1 - \frac{1}{2\|M\|^2}\tr\left[U^{(\nu)^T} \left(Q^{(\nu)}_\text{b} + Q^{(\nu)}_\text{f}\right) U^{(\nu)}\right]  ,
\label{symerr}
\end{align}
where we also used $\tr Q^{(\nu)}_\text{b} = \tr Q^{(\nu)}_\text{f} = \|M\|^2$.
We define the SuperQ matrix for direction $\nu$ as
\begin{align}
Q^{(\nu)}_\text{S} = Q^{(\nu)}_\text{b} + Q^{(\nu)}_\text{f} ,
\label{QS}
\end{align}
which is symmetric and positive semi-definite as it is a sum of two symmetric positive semi-definite matrices. Therefore, if we diagonalize the SuperQ matrix and truncate the eigenvector matrix, retaining the eigenvectors corresponding to the $K_\nu$ largest eigenvalues, then this semi-orthogonal frame $U^{(\nu)}$ minimizes the truncation error \eqref{symerr} on $Q^{(\nu)}_\text{S}$. This SuperQ procedure is repeated on all $d$ directions to determine all frames $U^{(\nu)}$, which can then be used to approximate $M$ and to construct the corresponding core tensor $S$, see \eqref{core-d}.

The SuperQ method is computationally efficient since it only requires a single eigenvalue decomposition for each pair of backward and forward fat modes, while the standard HOTRG procedure \cite{Xie_2012} performs separate decompositions on these modes.

When applying the SuperQ method to HOTRG, where $M$ is a contraction $T \star_\nu T$ along one of the directions, only $2d-2$ modes will actually be truncated, as the backward and forward modes for the contracted direction need not be truncated.

In analogy to the interlaced HOSVD approximation, see Sec.~\ref{Sec:HOSVD}, we can also define an interlaced version of the SuperQ method where we determine the frames $U^{(\nu)}$ by applying the method to an intermediate core tensor, which gets updated direction-by-direction (starting from $S=M$) by truncating the respective backward and forward mode each time a frame has been computed.
This interlaced SuperQ method is also of interest in the light of the iterative procedure derived in the next section.

\section{Optimized frames with iterative SuperQ}
\label{Sec:BFHOOI}

Although the HOSVD method, see Sec.~\ref{Sec:HOSVD}, typically yields a good tensor approximation $\hat M$ \cite{DeLathauwer2000}, the best-possible one, which minimizes $\|M-\hat M\|^2$, can be constructed with an iterative procedure called higher order orthogonal iteration (HOOI) method \cite{DeLathauwer2000a}.

According to the discussion of the backward-forward symmetric truncation in Sec.\ \ref{Sec:BFcond}, it is clear that HOOI is itself not applicable in a tensor network approach to statistical physics, because the backward and forward modes should be truncated with the same semi-orthogonal frame for each direction, while HOOI very generically generates different frames for all modes. Below we present the iterative SuperQ (ISQ) method, which is inspired by the original HOOI procedure but imposes the requirement that the same frame has to be used to truncate the backward and forward modes of each direction.

As in Sec.~\ref{Sec:SuperQ}, we consider a tensor $M$ with $d$ pairs of backward and forward modes, i.e., the tensor is of order $2d$ with dimensions $N_1\times N_1\cdots\times N_{d}\times N_{d}$, which will be truncated to dimensions $K_1 \times K_1 \times \cdots \times K_d \times K_d$, see \eqref{core-d}. 
Our aim is to minimize the squared Frobenius norm \eqref{normapprox}
\begin{align}
\|M-\hat M\|^2 = \|M\|^2-\|S\|^2 ,
\label{normbis}
\end{align}
over all semi-orthogonal $N_\nu\times K_\nu$ frames $U^{(\nu)}$, $\nu=1\ldots d$, for fixed $K_\nu$, with the additional condition that the backward and forward modes for each direction $\nu$ are truncated with the same frame $U^{(\nu)}$. 

The semi-orthogonality of the frames is imposed explicitly by orthonormality conditions for the column vectors of $U^{(\nu)}$,
\begin{align}
\sum_{i=1}^{N_\nu} U^{(\nu)}_{ia} U^{(\nu)}_{ib} = \delta_{ab} , \qquad \text{for } 1 \leq a,b \leq K_\nu \text{ and } 1 \leq \nu \leq d,
\label{condition}
\end{align}
in the constrained minimization of \eqref{normbis}. This leads to the cost function
\begin{align}
g = \|S\|^2 + C 
= \| U^{(1)^T} \odot_1 U^{(1)^T} \odot_2 \cdots U^{(d)^T} \odot_{2d-1} U^{(d)^T} \odot_{2d} M \|^2 
+ \sum_{\nu=1}^d \tr\left[\Lambda^{(\nu)}\left(\mathbbm 1 - U^{(\nu)^T}U^{(\nu)}\right)\right]
,
\label{costfun}
\end{align}
with matrices $\Lambda^{(\nu)}$ containing the Langrange multipliers. The orthonormalization conditions are symmetric under the exchange $a \leftrightarrow b$, and so $\Lambda^{(\nu)}$ will be symmetric too. If we diagonalize $\Lambda^{(\nu)}_\text{diag}=O^{(\nu)^T}\Lambda^{(\nu)} O^{(\nu)}$, redefine $U^{(\nu)}O^{(\nu)} \to U^{(\nu)}$ and $\Lambda^{(\nu)}_\text{diag} \to \Lambda^{(\nu)}$, and use the orthogonality of $O^{(\nu)}$, then Eq.\eqref{costfun} remains unaltered albeit now with diagonal $\Lambda^{(\nu)}=\text{diag}\big(\lambda^{(\nu)}_1,\ldots,\lambda^{(\nu)}_{K_\nu}\big)$, and this without loss of generality.
Written out in components this is
\begin{align}
g 
&= \sum_{\{a,b\}} S_{a_1 b_1\cdots a_d b_d}^2 
+ \sum_{\nu=1}^d \sum_{c=1}^{K_\nu} \lambda^{(\nu)}_{c} \left(1 - \sum_{k=1}^{N_\nu} U^{(\nu)^2}_{kc} \right) 
\notag\\
&= \sum_{\{a,b\}} \left(\sum_{\{i,j\}} \bigg[\prod_{\mu=1}^d U^{(\mu)}_{i_\mu a_\mu} U^{(\mu)}_{j_\mu b_\mu}\bigg] M_{i_1 j_1\cdots i_d j_d}\right)^2 
+ \sum_{\nu=1}^d \sum_{c=1}^{K_\nu} \lambda^{(\nu)}_{c} \left(1 - \sum_{k=1}^{N_\nu} U^{(\nu)^2}_{kc} \right) 
.
\end{align}
For a constrained maximum of $\|S\|^2$, the partial derivative of $g$ with respect to the $(k,c)$-entry of the $\nu$-th orthogonal frame has to satisfy
\begin{align}
0 = \frac{\partial g}{\partial U^{(\nu)}_{kc}}
&= \sum_{\{a,b\}} 2 S_{a_1 b_1\cdots a_d b_d} \frac{\partial}{\partial U^{(\nu)}_{kc}} 
\left(\sum_{\{i,j\}} \bigg[\prod_{\mu=1}^d U^{(\mu)}_{i_\mu a_\mu} U^{(\mu)}_{j_\mu b_\mu}\bigg] M_{i_1 j_1\cdots i_d j_d}\right) 
- 2 \lambda^{(\nu)}_c U^{(\nu)}_{kc} ,
\end{align}
for $1\leq k \leq N_\nu$, $1 \leq c \leq K_\nu$ and $1 \leq \nu \leq d$.
Note that the same frame $U^{(\nu)}$ appears twice in $S$, as it is used to truncate the modes in the backward and forward $\nu$ direction. We therefore obtain
\begin{align}
\sum_{\{a,b\}} S_{a_1 b_1 \cdots a_d b_d}
&\left(
\sum_{\{i,j\}} \delta_{k,i_\nu} \delta_{c,a_\nu} \bigg[\prod_{\mu\neq\nu} U^{(\mu)}_{i_\mu a_\mu} U^{(\mu)}_{j_\mu b_\mu}\bigg] U^{(\nu)}_{j_{\nu}b_{\nu}} M_{i_1 j_1\cdots i_n j_n} \right.\notag\\
 &\left. + \sum_{\{i,j\}} \delta_{k,j_{\nu}} \delta_{c,b_{\nu}} \bigg[\prod_{\mu\neq\nu} U^{(\mu)}_{i_\mu a_\mu} U^{(\mu)}_{j_\mu b_\mu}\bigg] U^{(\nu)}_{i_\nu a_\nu} M_{i_1 j_1\cdots i_n j_n}\right) 
= \lambda^{(\nu)}_c U^{(\nu)}_{kc} 
.
\end{align}
After eliminating the Kronecker deltas we get
\begin{align}
&\sum_{\{a,b\}\setminus a_{\nu}} S_{a_1 b_1 \cdots c b_{\nu} \cdots a_d b_d} 
\sum_{\{i,j\}\setminus i_{\nu}} \bigg[\prod_{\mu\neq\nu} U^{(\mu)}_{i_\mu a_\mu} U^{(\mu)}_{j_\mu b_\mu}\bigg] U^{(\nu)}_{j_{\nu}b_{\nu}} M_{i_1 j_1\cdots k j_{\nu} \cdots i_d j_d} \notag\\
 & + \sum_{\{a,b\}\setminus b_{\nu}} S_{a_1 b_1 \cdots a_{\nu} c \cdots a_d b_d} \sum_{\{i,j\}\setminus j_{\nu}}  \bigg[\prod_{\mu\neq\nu} U^{(\mu)}_{i_\mu a_\mu} U^{(\mu)}_{j_\mu b_\mu}\bigg] U^{(\nu)}_{i_\nu a_\nu} M_{i_1 j_1\cdots i_{\nu}k \cdots i_{d}j_d}
= \lambda^{(\nu)}_c U^{(\nu)}_{kc} .
\label{withoutBandF}
\end{align}
Let us define the unfolding matrices $B^{(\nu)}$ and $F^{(\nu)}$ with dimensions $N_\nu \times (K_1^2 \cdots K_{\nu-1}^2 K_\nu K_{\nu+1}^2\cdots K_d^2)$, where all directions of $M$ are truncated, except for the backward-$\nu$ mode for $B^{(\nu)}$, and the forward-$\nu$ mode for $F^{(\nu)}$, and the unfolding is performed with respect to the untruncated mode,
\begin{align}
B^{(\nu)}_{i_\nu\,,\,(\{a,b\}\setminus a_\nu)} &= 
\sum_{\{i,j\}\setminus i_{\nu}} \bigg[\prod_{\mu\neq\nu} U^{(\mu)}_{i_\mu a_\mu} U^{(\mu)}_{j_\mu b_\mu}\bigg] U^{(\nu)}_{j_{\nu}b_{\nu}}  M_{i_1 j_1\cdots \cdots i_d j_d}  \label{B-unfold}\\
F^{(\nu)}_{j_\nu\,,\,(\{a,b\}\setminus b_\nu)} &= 
\sum_{\{i,j\}\setminus j_{\nu}} \bigg[\prod_{\mu\neq\nu} U^{(\mu)}_{i_\mu a_\mu} U^{(\mu)}_{j_\mu b_\mu}\bigg] U^{(\nu)}_{i_\nu a_\nu} M_{i_1 j_1\cdots i_d j_d} \,,
\label{F-unfold}
\end{align}
where we used the notation introduced in \eqref{unfold} for the matrix indices.
The core tensor can also be written in terms of $B^{(\nu)}$ and $F^{(\nu)}$ by truncating the remaining untruncated index:
\begin{align}
S_{a_1 b_1 \cdots a_d b_{d}}  
&= \sum_{i_\nu} B^{(\nu)}_{i_\nu\,,\,(\{a,b\}\setminus a_\nu)} U^{(\nu)}_{i_\nu a_\nu}
= \sum_{j_\nu} F^{(\nu)}_{j_\nu\,,\,(\{a,b\}\setminus b_\nu)} U^{(\nu)}_{j_\nu b_\nu}.
\label{S(BF)}
\end{align}
After substituting $B^{(\nu)}$ and $F^{(\nu)}$ in \eqref{withoutBandF} we obtain
\begin{align}
&\sum_{\{a,b\}\setminus{a_\nu}} \sum_{i_\nu}  
B^{(\nu)}_{k\,,\,(\{a,b\}\setminus a_\nu)} B^{(\nu)}_{i_\nu\,,\,(\{a,b\}\setminus a_\nu)}  U^{(\nu)}_{i_\nu c} 
+ \sum_{\{a,b\}\setminus{b_\nu}} \sum_{j_\nu} F^{(\nu)}_{k\,,\,(\{a,b\}\setminus b_\nu)} F^{(\nu)}_{j_\nu\,,\,(\{a,b\}\setminus b_\nu)} U^{(\nu)}_{j_\nu c}
= \lambda^{(\nu)}_c U^{(\nu)}_{k c} ,
\end{align}
or
\begin{align}
\sum_{i_\nu} \left[B^{(\nu)} B^{(\nu)^T}\right]_{k i_\nu} U^{(\nu)}_{i_\nu c} 
+ \sum_{j_\nu} \left[F^{(\nu)} F^{(\nu)^T}\right]_{k j_\nu} U^{(\nu)}_{j_\nu c}
= \lambda^{(\nu)}_c U^{(\nu)}_{kc}  .
\label{entry-like}
\end{align}
If we introduce the positive semi-definite matrices
\begin{align}
Q^{(\nu)}(U^{(1)},\ldots,U^{(d)}) = B^{(\nu)} B^{(\nu)^T} + F^{(\nu)} F^{(\nu)^T}, \qquad\qquad \nu=1,\ldots, d,
\label{Qr}
\end{align}
we can identify \eqref{entry-like} as a coupled nonlinear eigenvalue problem (which is nonlinear in the eigenvectors) 
\begin{align}
\phantom{\qquad\qquad \nu=1\ldots d,}
Q^{(\nu)}(U^{(1)},\ldots,U^{(d)}) \, U^{(\nu)} &= U^{(\nu)}\Lambda^{(\nu)}  , \qquad\qquad \nu=1,\ldots, d,
\label{bfhooi}
\end{align}
for the semi-orthogonal frames $U^{(\nu)}$ of dimension $N_\nu \times K_\nu$ and the $K_\nu$-dimensional diagonal matrices $\Lambda^{(\nu)}$.
Note that all frames $U^{(\mu)}$, $\mu=1\ldots d$, appear in $Q^{(\nu)}$ (as projections $P^{(\mu)}=U^{(\mu)}U^{(\mu)^T}$). This is even true for the direction $\nu$ itself, as its projector is applied to the forward mode in $B^{(\nu)} B^{(\nu)^T}$ and to the backward mode in $F^{(\nu)} F^{(\nu)^T}$. 
Therefore the $\nu$-th equation is cubic in $U^{(\nu)}$ and quartic in all other $U^{(\mu)}$, $\mu\neq\nu$.
It is crucial to keep in mind that the self-consistent solutions to \eqref{bfhooi}, which we are looking for, are required to be semi-orthogonal matrices with $K_\nu \leq N_\nu$ columns, in order to satisfy the constraints \eqref{condition}.

It is useful to note that if we replace the matrices $Q^{(\nu)}$ in \eqref{bfhooi} by fixed matrices $\Qhat^{(\nu)}$, the $d$ matrix equations decouple and each one of them is a linear eigenvalue equation for the frame $U^{(\nu)}$. The solutions of these linearized equations are however in general no solution of the original nonlinear equations \eqref{bfhooi}. On the other hand, if \eqref{bfhooi} is satisfied, then $U^{(\nu)}$ is a solution of the linear eigenvalue problem for the specific matrix $\Qhat^{(\nu)}=Q^{(\nu)}(U^{(1)},\ldots,U^{(d)})$.

Therefore, we propose to solve the coupled system of equations using an iterative procedure,
where at each iteration step, $\Qhat^{(\nu)}$ is computed with \eqref{Qr} using the current frames $U^{(\mu)}$, $\mu=1,\ldots,d$, and the 
eigenvalue problem
\begin{align}
\Qhat^{(\nu)}(U^{(1)},\ldots,U^{(d)})  \, u^{(\nu)} = \lambda^{(\nu)} u^{(\nu)} ,
\label{itproc}
\end{align}
is solved for the eigenvalues $\lambda^{(\nu)}$ and eigenvectors $u^{(\nu)}$ of the positive semi-definite matrix $\Qhat^{(\nu)}$. We then take the normalized eigenvectors corresponding to the $K_\nu$ largest eigenvalues to form a new frame $U^{(\nu)}$. Note that the eigenvectors of the symmetric matrix $\Qhat^{(\nu)}$ are orthogonal and therefore the constraint \eqref{condition} is automatically satisfied for $U^{(\nu)}$.

This strategy can be motivated in the following way. We are looking for the solution of \eqref{bfhooi} which maximizes $\|S\|^2$. Using the expression \eqref{S(BF)} for the core tensor $S$ in terms of the unfoldings $B^{(\nu)}$ and $F^{(\nu)}$, the squared norm in the cost function \eqref{costfun} can be written as
\begin{align}
\|S\|^2 = \frac12 \left(\tr[U^{(\nu)^T} B^{(\nu)} B^{(\nu)^T} U^{(\nu)}] 
+ \tr[U^{(\nu)^T} F^{(\nu)} F^{(\nu)^T} U^{(\nu)}] \right)
= \frac12 \tr[U^{(\nu)^T} Q^{(\nu)} U^{(\nu)}] ,
\label{normS}
\end{align}
which explicitly contains the matrix $Q^{(\nu)}\equiv Q^{(\nu)}(U^{(1)},\ldots,U^{(d)})$ defined in \eqref{Qr} (note that this equation yields the same $\|S\|^2$ for each $\nu=1,\ldots,d$). It is straightforward to show that, at each iteration step, the solution obtained from the linearization \eqref{itproc} corresponds to the constrained optimization of the  approximation
\begin{align}
\|S\|^2 \approx \frac12 \tr[U^{(\nu)^T} \Qhat^{(\nu)} U^{(\nu)}]
\label{normSapprox}
\end{align}
to \eqref{normS}, where $\Qhat^{(\nu)}$ is fixed and computed with the most recent frames $U^{(\mu)}$, $\mu=1,\ldots,d$.  As this approximation is quadratic in the new frame $U^{(\nu)}$, its maximum will be given by the eigenvectors corresponding to the $K_\nu$ largest  eigenvalues of the symmetric, positive semi-definite matrix $\Qhat^{(\nu)}$.

The iterative procedure can be interpreted as an iterative SuperQ method, where at each step all modes of $M$ are truncated using the last known frames, except for a backward mode in $B$ and the corresponding forward mode in $F$.

During the iterative procedure we cycle through the $d$ dimensions $\nu$ and determine a new frame $U^{(\nu)}$ at each step using \eqref{itproc}. Then, we repeat these $d$ iteration steps until all frames have converged. In practice we observed that the first iteration for each direction is the most important one, and further iterations of the same direction only give small corrections. 

An alternative procedure would be to iterate \eqref{itproc} for a single $U^{(\nu)}$ (keeping all other $U^{(\mu)}$ fixed) until convergence has been reached (reevaluating $\Qhat^{(\nu)}$ with the most recent $U^{(\nu)}$ at every step), and then go on to the next frame. Once all frames have been iterated, this whole procedure is repeated until all frames converge together. 
Note that the inner iterations are computationally cheap, as all frames but one are kept fixed and all matrix-tensor multiplications in $\Qhat$ and $\|S\|^2$ involving these fixed frames have to be computed only once.
However, this procedure does not seem to give an overall faster convergence.

A natural choice for the starting frames $U^{(\nu)}$, $\nu=1\ldots d$, in the iterative procedure are the frames obtained from the interlaced SuperQ method, see Sec.~\ref{Sec:SuperQ}.

Note that the iterative procedure is not guaranteed to converge, and even when it does, the solution is not necessarily the global maximum. This can be improved upon by tuning the starting frames or by applying an under-relaxation procedure
to the intermediate $Q$ matrix. In this procedure, we replace $\Qhat^{(\nu)}$ in \eqref{itproc} by
\begin{align}
\overline Q^{(\nu)} = \omega \, \Qhat^{(\nu)} + (1-\omega) \, \Qhat^{(\nu)}_\text{prev} \,,
\label{relax}
\end{align}
where $\Qhat^{(\nu)}_\text{prev}$ was used to obtain the previous $U^{(\nu)}$ in the iterative procedure.
The relaxation procedure can be used to optimize $\|S\|^2$ by tuning the local parameter $0 \leq \omega \leq 1$. We observed that a coarse tuning of $\omega$ is sufficient to improve the overall convergence of the iterative procedure.

When applying the ISQ method to HOTRG, where $M$ is a contraction $T \star_\nu T$ along one of the directions, only $2d-2$ of $2d$ modes will actually be truncated, as the backward and forward modes for the contracted direction are left unchanged.
For the two-dimensional case, where only one frame has to be determined after each contraction, an alternative method to optimize the truncation using a linearization was proposed in the \textit{projective truncation} of Ref.\ \cite{Evenbly2015}.

\section{Examples}

In the following, we illustrate the effect of the SuperQ and ISQ methods. For various random tensors $A$, we compute core tensors $S$ using the standard HOSVD approximation \cite{DeLathauwer2000} and the best possible approximation of a given multi-rank with the HOOI method \cite{DeLathauwer2000a}. These results are compared with the following backward-forward symmetric approximations: the method proposed by Xie et al.\ (used in standard HOTRG \cite{Xie_2012}), the SuperQ approximation of Sec.~\ref{Sec:SuperQ}, and the ISQ approximation of Sec.~\ref{Sec:BFHOOI}.
The relative error for each of these approximations is given by
\begin{align}
\epsilon = \sqrt{1-\frac{\|S\|^2}{\|A\|^2}} .
\end{align}
The comparison of the different methods will be illustrated by plotting $\epsilon-\epsilon_\text{hooi}$ in the figures below.

\begin{figure}
\centering
\includegraphics[width=0.49\textwidth]{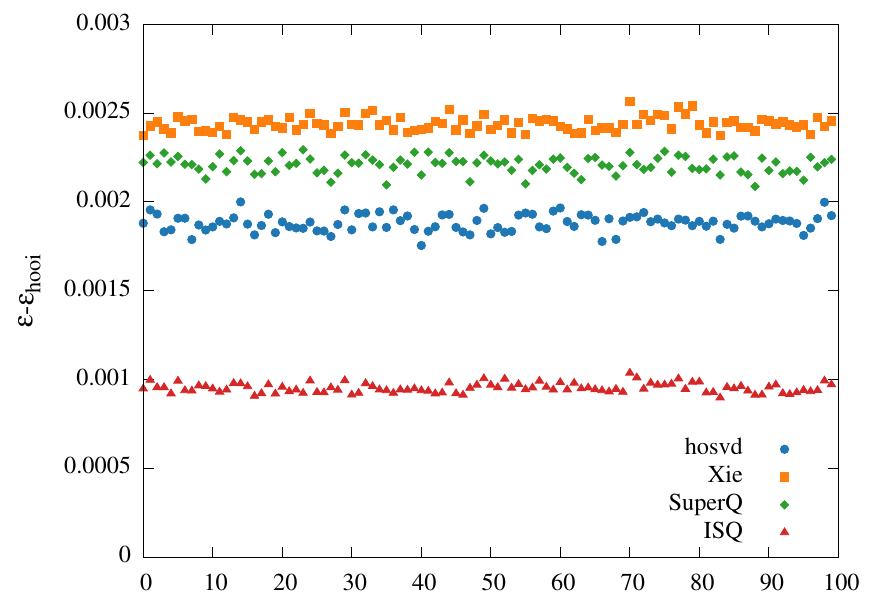}
\includegraphics[width=0.49\textwidth]{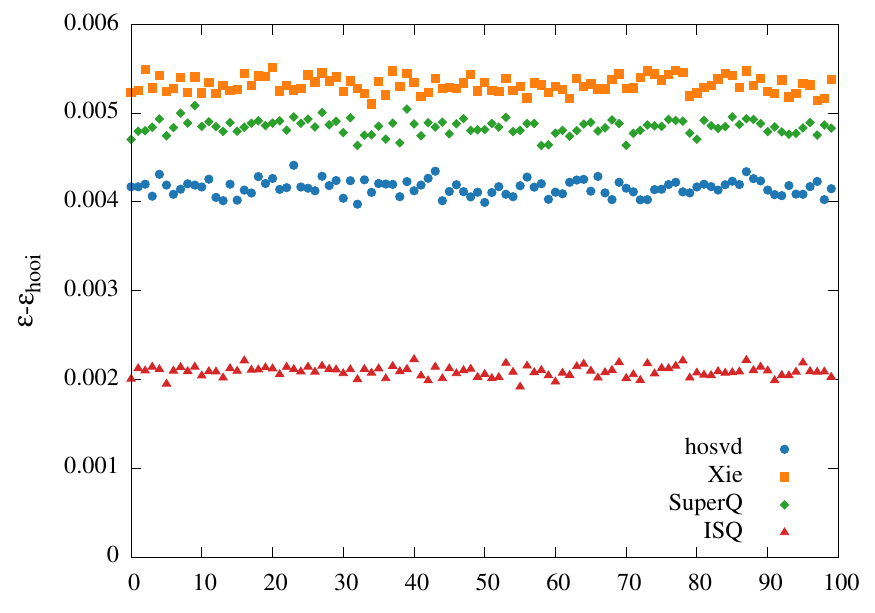}
\caption{Comparison of $\epsilon-\epsilon_\text{hooi}$ for approximations of 100 independent random tensors of dimension $10\times10\times100\times100$ truncated to dimension $10\times10\times10\times10$ using HOSVD, the Xie method, SuperQ and ISQ. The tensor elements are chosen randomly from a uniform distribution over [0,1] (left plot) and from a Gaussian distribution $N(0;1)$ (right plot). The horizontal axis represents different random tensors.}
\label{fig:1010100100}
\end{figure}

\begin{figure}
\centering
\includegraphics[width=0.49\textwidth]{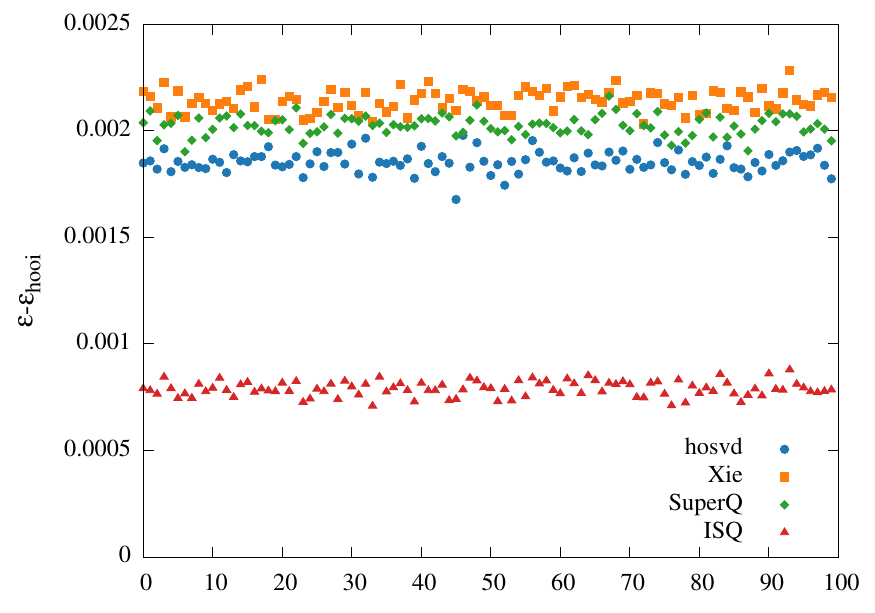}
\includegraphics[width=0.49\textwidth]{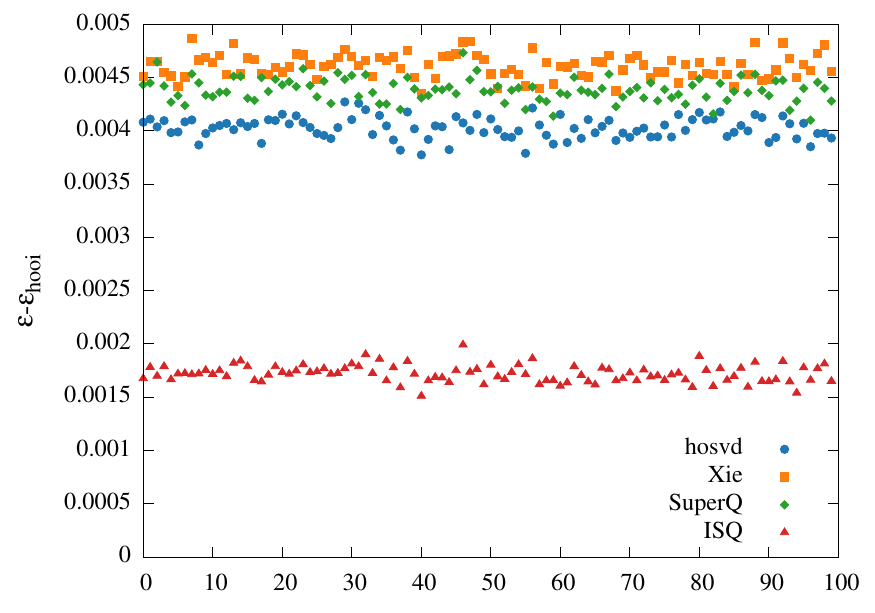}
\caption{Comparison of $\epsilon-\epsilon_\text{hooi}$ for random tensors of dimension $30\times30\times30\times30$ truncated to dimension  $10\times10\times10\times10$, using the same approximation methods and the same probability distributions for the tensor elements as in Fig.~\ref{fig:1010100100}. }
\label{fig:30303030}
\end{figure}

In a first example we consider random tensors $A$ of order 4 with dimension $10\times10\times100\times100$, whose rank is reduced by truncating the last two indices to dimension 10. The results shown in Fig.~\ref{fig:1010100100} were computed for initial tensors filled with uniformly distributed elements in $[0,1]$ (left panel) and normally distributed elements with mean $\mu=0$ and standard deviation $\sigma=1$ (right panel). In another example, shown in Fig.~\ref{fig:30303030}, all modes of random $30\times30\times30\times30$ tensors are reduced to dimension 10. Again the random tensors are filled with elements from a uniform distribution (left) and a normal distribution (right).

In all examples, the hierarchy between the approximations is the same. In decreasing order of accuracy we find: HOOI, ISQ, HOSVD, SuperQ, and finally the Xie-method. We notice that, as we suggested in the derivation of Sec.~\ref{Sec:SuperQ}, the SuperQ method performs better than the Xie-method. Both of them are superseded by HOSVD, which is logical as the latter does not have to satisfy the additional backward-forward symmetry constraint. However, a somewhat unexpected result is that in all examples, the ISQ method performs better than the standard HOSVD approximation, even though the former does satisfy the additional backward-forward symmetry constraint. In all cases HOOI performs best, as it is the best possible approximation of the given multi-rank.

\section{Conclusions}
\label{Sec:conclusions}

In this paper we consider the reduction of the local truncation error in a single blocking step of the HOTRG procedure. 
We have discussed in detail the constraints imposed on the semi-orthogonal truncation frames in the HOTRG algorithm, where the backward and forward modes for each direction have to be projected on the same lower-dimensional subspace at each blocking step.
We first introduced the SuperQ method, which minimizes a combined error on the backward and forward unfoldings for each individual direction.
The method is computationally more efficient and generically yields a reduced local truncation error when compared to the original HOTRG truncation.

As a further improvement, we presented the iterative SuperQ method, where we formulate a constrained minimization problem, which leads to equations that have to be satisfied by the semi-orthogonal truncation frames in order to minimize the error on the lower-rank tensor approximation, while satisfying the backward-forward symmetry constraints. The method is inspired by the HOOI method, with the additional requirement that the same frames are used on the backward and forward modes of each direction. The equations form a coupled nonlinear eigenvalue problem, which we propose to solve using an iterative procedure, where decoupled linear eigenvalue problems are solved at each iteration step. Computing the optimal backward-forward symmetric truncation frames with the ISQ method is more expensive than the truncation applied in the original HOTRG method, as each iteration step requires new eigenvalue decompositions. Therefore, in practice, we generally use the SuperQ truncation in the HOTRG blocking procedure, since it provides the best trade-off between computational cost and truncation accuracy.


\bibliographystyle{elsarticle-num.bst}
\bibliography{biblio} 

\end{document}